\RequirePackage{fix-cm}
\documentclass[smallextended]{svjour3}       
\smartqed  
\usepackage{graphicx}
\begin{document}

\title{A Core-envelope Analytic Model for the Vela Pulsar
}


\author{P S Negi       
}


\institute{Department of Physics, \at
              Kumaun University, Nainital\\
              \email{psneginainital63@gmail.com}           
}

\date{Received: date / Accepted: date}

\maketitle

\begin{abstract}
The core-envelope models presented in Negi et al. \cite{Ref1}; Negi \cite {Ref2}, corresponding to the values of compactness parameter, $u \equiv M/a$ = 0.30 and 0.25 (mass to size ratio in geometrized units) have been studied under slow rotation. It is seen that these models are capable of explaining all the observational values of glitch healing parameter, $G_h = I_{\rm core}/I_{\rm total} < 0.55$ \cite {Ref3} ( $G_h$ represents the fractional moment of inertia of the core component  in the starquake mechanism of glitch generation) for the Vela pulsar. The models yield the maximum values of mass, $M$, surface redshift, $z_a$, and the moment of inertia, $I_{\rm Vela}$ for the Vela pulsar in the range $M = 3.079M_\odot - 2.263M_\odot$; $z_a = 0.581 - 0.414$ and $I_{\rm Vela,45} =6.9 - 3.5$ (where $I_{45}=I/10^{45}\rm g{cm}^2$) respectively for the values of $u = $ 0.30 and 0.25 and for an assigned value of the surface density, $E_a = 2\times 10^{14}\rm g{cm}^{-3}$ \cite{Ref4}. The values of masses lower than the above mentioned values ( so called the realistic mass range, $M = 1.4\pm0.2 M_\odot$, in the literature) but significantly higher than that of the unrealistic mass range $M  \leq 0.5M_\odot$ (obtained for the Vela pulsar in the literature on the basis of parametrized neutron star (NS) models based on equations of state (EOSs) of dense nuclear matter \cite {Ref3}) and other parameters may be obtained likewise for the above mentioned  range of the  values of $G_h$ corresponding to the values of $u < 0.25$. The models are found to be causally consistent, gravitational bound  and pulsationally stable. The upper bound on neutron star (NS) mass obtained in this study which is applicable  for the Vela pulsar, in fact,  corresponds to the mean value of the  upper bound on NS mass obtained in the classical result by Rhoades \& Ruffini \cite {Ref5} and that obtained  on the basis of modern EOSs for neutron star matter by Kalogera \& Byam \cite {Ref6} and is in a good agreement with the most recent theoretical estimate made by Sotani \cite {Ref7} .
\end{abstract}

\section{Introduction}
\label{intro}
The observational data on glitch for the Crab and the Vela pulsar have been extensively studied in the literature (see, Crawford \&  Demiansky \cite{Ref3}; and references therein) which provide best tool for testing different glitch models (mechanism for glitch generation in pulsars). Since the glitch models are based on the internal structure of neutron stars, thus they can provide important information in this regard. In the starquake glitch  model \cite{Ref8}, \cite{Ref9} a NS is considered as a two-component structure consisting a superfluid core which consists most of the mass of the NS, surrounded by a rigid crust which contains only a few percent of the total mass. As a result of considering the thin crust in the conventional NS models, the larger values of glitch healing parameter (weighted mean value), $G_h = I_{\rm core}/I_{\rm total} \geq 0.7$ obtained for the Crab pulsar are easily satisfied by most of the conventional NS models in the realistic mass range ($M = 1.4\pm0.2 M_\odot$)[\cite {Ref3}; and references therein]. Whereas the much lower values of glitch healing parameter (weighted mean value), $G_h \leq 0.2 $, obtained for the Vela pulsar could not be satisfied on the basis of starquake model unless the Vela mass falls  in the unrealistic low mass range, $M \leq 0.5M_\odot$ \cite{Ref3}. Furthermore, the estimation of core-crust boundary in the conventional NS models in somewhat arbitrary in the sense that one can freely choose somewhat lower value of the boundary (which would increase the core size) to get higher values of the glitch healing parameter, $G_h$, (see, e.g. \cite {Ref10}, \cite {Ref11}). In order to avoid the said  arbitrariness in the conventional NS models, earlier the author\cite{Ref12} had used a certain `criterion' and obtained the core-envelope (instead of the term `crust' used in the conventional NS models, the term `envelope ' is used here which includes the crust plus other interior part of the star right upto the superfluid core) boundary of the NS models consisting a core of  stiffest matter (adiabatic sound speed, $dP/dE$ = 1, $P$ and $E$ are the pressure and energy density) and the envelope is described by the EOS of a classical polytrope. This model has been successful to explain the glitch healing parameter (weighted mean) of the Vela pulsar in the range $0 < G_h \leq 0.197$ corresponding to the realistic mass range $1.758 \leq M \leq 2.2M_\odot$ for the Vela pulsar.

Besides the two recent precisely measured larger pulsar masses about  $2M_\odot$ (\cite{Ref13}, \cite{Ref14}), the theoretical estimation indicates that the upper limit on NS mass may increase significantly from the said measured values and  lie between $2 - 3.05M_\odot$  \cite{Ref7}. In recent studies, Zhou et al \cite {Ref15} discussed the starquake model for the Vela pulsar on the basis of a solid quark star model. Lai et al \cite{Ref16} have proposed a strangeon star model  (i.e., the solidification of the star takes place during cooling) and studied the behaviors of glitches (as result of starquake) without significant energy release, including the Crab and the Vela pulsars. By using NS observations, Steiner et al\cite {Ref17} obtained a crustal fraction of the moment of inertia as large as 10\% for a mass $M = 1.4M_\odot$ to explain the glitches in the Vela pulsar even with a large amount of superfluid `entrainment'(\cite {Ref18}; \cite {Ref19}). Delsate et al\cite {Ref20} have calculated the crustal moment of inertia of glitching pulsars for different unified dense matter EOSs in order to explain the large glitches observed  in the case of the Vela pulsar.

 The present study continues to deal with the construction of such NS models which can satisfy the all observational values  (not only the weighted mean values considered in the previous study) of the glitch healing parameter, $0 < G_h < 0.55$, obtained for the Vela pulsar together with the property that they  may correspond to the `realistic' mass range of NSs  which ends at the value of maximum mass. The maximum value of  NS mass obtained in this study may be considered identical to those obtained  on the basis of modern EOSs of dense nuclear matter available the literature (\cite {Ref6};  \cite {Ref7}; and references therein). Since various observational studies and their explanation favour smooth density variation inside the NS structure (\cite{Ref21}, \cite{Ref22}, \cite{Ref23}, \cite{Ref24}), therefore it is expected that for realistic NS models the (energy) density in the core region varies smoothly as compared to that of the  outer region (envelope) where the variation of density becomes relatively faster. We have, therefore, considered in the present study, the core-envelope models presented in Negi et al. \cite{Ref1}; Negi \cite{Ref2} in which the density in the core is governed by the smooth variation of density (Tolman's type VII solution) and the density in the envelope is described by the Tolman's type V solution. The solution in the core fulfills various criteria of physically realistic structures ( \cite {Ref25}, \cite{Ref26}) whereas the solution considered in the envelope is found to be entirely  analogous to the EOS of Wringa et al.\cite{Ref27}(WFF) which is widely used in the envelope region of various NS models discussed in the literature \cite{Ref6} (the analogy of Tolman's V solution used in the envelope of the present study with that of the EOS WFF is explained in Sec 3). Thus, like various NS models based on EOS discussed in the literature, the models presented here, on the basis of exact analytical solutions of Einstein's field equations, are causally consistent, gravitationally bound  and pulsationally stable for the allowed values of the compactness parameter $u \leq 0.30$ ( Negi \cite{Ref2}). The distinctive feature of the models considered in the present study lies in the fact that unlike various NS models available in the literature in which the choice of the core-envelope boundary bears some arbitrariness  \cite {Ref9}, \cite {Ref10}, the boundary of the core-envelope models presented in \cite{Ref1}; Negi \cite{Ref2} has been  obtained by an appropriate matching of  all the four variables viz. pressure ($P$), energy density ($E$) and both of the metric parameters ($\nu$ and $\lambda$) at the core envelope boundary. The model yields larger values of radius for a given mass as compared to that of the previous study of the author ( Negi \cite{Ref11}) and may provide larger values of moment of inertia for the configuration.

The study carried out by Negi et al. \cite{Ref1} deals with the construction of a core-envelope model of static and spherical mass distribution characterized
 by exact solutions of Einstein's field equations. The core of the model is described by Tolman's VII  solution matched smoothly at the core-boundary. The region of the envelope is
described by Tolman's V solution which is finally matched to the vacuum Schwarzschild solution.  The complete solutions with appropriate references for both the regions (the core and the envelope) are available in  Negi et al. \cite{Ref1}. However,  Negi  \cite{Ref2} has re-investigated these models by removing an error in the models of  Negi et al. \cite{Ref1} by  rewriting equations (19) - (22 )[as eqs. (1) - (5) and expression for $w_b$ as eq. (6)] in Sec. 2 of the paper  Negi \cite{Ref2} with the single replacement in the symbol $`t' \equiv `Q'$ which was assigned as a compressibility parameter in Tolman's  VII solution $ (x = r^2/K^2 = r^2/a^2t)$ discussed in  Negi et al. \cite{Ref1} and recomputed the parameters. The modified results with some other important properties of the models (adiabatic sound speed $({\rm d}P/{\rm d}E)_0$ at the centre,  gravitational binding energy and the pulsational stability under small radial perturbations) which were not discussed in the paper of  Negi et al. \cite{Ref1} are also included in  Negi \cite{Ref2}.

 Sect.2 of the present study deals with the relevant equations governing slow rotation of the core-envelope models presented in Negi et al. \cite{Ref1}; Negi \cite{Ref2} . Results of the calculations and an application of the models to the Vela pulsar are presented in Sect.3.  Sect 4  summarizes  the main  results of the present study.

\section{Equations Governing Slow Rotation of Spherical Configuration}

The metric corresponding to a static and spherically symmetric mass distribution is given by
\begin{equation}
ds^2 = e^\nu dt^2 - e^\lambda dr^2 - r^2(d\theta^2 +{\rm sin}^2\theta d\phi^2)
\end{equation}
where $G = c = 1$ (we are using geometrized units) and $\nu$ and $\lambda$ are functions of $`r'$ alone.
The relevant equations governing the core ($0\leq r \leq b$) and  the envelope ($b\leq r \leq a$) are described respectively by Tolman's type VII and V solutions of the metric (Eq.1) and are available in 
 Negi et al. \cite{Ref1}; Negi \cite{Ref2}. However, some relations relevant to the present study are redefined in the following:

$u \equiv M/a$ is called the compactness parameter which is defined as the mass to size (radius) ratio of entire configuration; where
 the mass, $M = \int_{0}^{a} 4\pi E r^2dr$; and $y = r/a$ is called the radial coordinate measured in units of configuration size. The parameter $n$ which appears in Tolma's type V solution is related with compactness ratio, $u$, as $u = n/(2n + 1)$.

$Q(\equiv K^2/a^2)$ is defined as the compressibility factor, $K$ is a constant appearing in the Tolman's type VII solution. The matching of various parameters at the core-envelope boundary yields the ratio $(b/a)$, thus $(b/a)$ represents the boundary, $y_b$, of the core-envelope models and $`b'$ represents the core radius.

For slowly rotating structures a perturbation solution for the metric (Eq. 1) yields (Borner \cite{Ref28}, Chandrasekhar and  Miller \cite{Ref29},
 Irvine \cite{Ref30})
\begin{equation}
(d/dr)[F(d\chi/dr)] = G \chi
\end{equation}
where
\begin{equation}
F = e^{-(\nu+\lambda)/2}r^4
\end{equation}
\begin{equation}
G = 16\pi(P + E) e^{(\lambda - \nu)/2}r^4
\end{equation}
\begin{equation}
\chi = \omega - \Omega
\end{equation}
 $\omega$ being the angular velocity of structure and $\Omega$ the drag of the local inertial tetrad (Hartle and Sharp \cite{Ref31},
 Hartle \cite{Ref32}); known as Lense-Thirring effect.

Substituting )$F(d\chi/dr)] = \phi$ in Eq. (2) we get
\begin{equation}
d\chi/dr = \phi/F
\end{equation}
and
\begin{equation}
d\phi/dr = G\chi
\end{equation}
Substituting $r = ay, \Delta r = a\Delta y$ in Eqs. (6) and (7) we get
$
d\chi/dy = (a\phi/F) = (a\phi/a^4f)
$
and
$(d\phi/dy) = G\chi a = a^2g\chi a
$
with $f =  e^{-(\nu+\lambda)/2}y^4$ and $g = 2(8\pi Pa^2 + 8\pi Ea^2)e^\lambda f$
or
\begin{equation}
[d/dy](\phi/a^3) =  g\chi
\end{equation}
and
\begin{equation}
d\chi/dy = (\phi/a^3)/f
\end{equation}
Substituting $\phi/a^3 = \psi$ in Eqs. (8) and (9) we have
\begin{equation}
d\psi/dy = g\chi
\end{equation}
\begin{equation}
d\chi/dy = \psi/f
\end{equation}
Eqs.(10) and (11) provide a set of two first order coupled differential equations which may be solved numerically by using the standard Runge - Kutta method with boundary conditions
\begin{equation}
\chi_{y=0} = 1; (d\chi/dy)_{y=0} = 0
\end{equation}
Integrating from the  centre ($y = 0$) to the surface ($y = 1$, i.e. $r = a$ and $P = 0$) of the configuration, we find that at the surface
\begin{equation}
\omega = \chi_a + (\phi_a/3a^3) = \chi_a + (\psi_a/3)
\end{equation}
Drag is given by the equation
\begin{equation}
\Omega = \omega - \chi; \rm {or} (\Omega/\omega) = 1 - (\chi/\omega)
\end{equation}
We define central drag as
\begin{equation}
 (\Omega/\omega)_0 = 1 - (1/\omega); \chi_0 = 1
\end{equation}
Thus the surface drag is given by
\begin{equation}
 (\Omega/\omega)_a = 1 - (\chi_a/\omega)
\end{equation}
The moment of Inertia, $I$, of the configuration is given by
\begin{equation}
  I = (\phi_a/6\omega) = (\psi_a a^3/6\omega)
\end{equation}



\section{Discussion of Results and Application of the Models to the Vela Pulsars}
The calculations are performed for the values of $u$ = 0.25 and 0.30 (corresponding to the $n$ values (1/2) and (3/4)) for assorted values of $Q$ in the range $0.001 - 1.3$ and an assigned value of the surface density, $E_a = 2 \times 10^{14}\rm {g cm}^{-3}$ (Brecher \& Caporaso \cite{Ref4}). The total size of the configuration depends only on $u$ value and turns out to be 13.369 km and 15.159 km for $u = 0.25$ and $u = 0.30$ respectively. The core size depends also on the value of $Q$ together with $u$. The masses of the models depend only on $u$  and have  the values 2.263$M_\odot$ and 3.079$M_\odot$ respectively for $u = 0.25$ and $u = 0.30$. The values of $(b/a)$, core mass, $M_b/M_\odot$, core size, $b$, boundary density, $E_b$, ratio of boundary density to surface density, $E_b/E_a$, and the glitch healing parameter, $G_h = I_b/I_a$ for the values of $u$ = 0.25 and 0.30 are shown in Table 1 and Table 2 respectively. It may be re-iterated that the maximum permissible value of $Q$ is obtained as 1.3 for which $(b/a) \approx 1$, i.e. the whole configuration pertains to just the core solution. For $Q = 0.001$ , we obtain $(b/a) \cong 0.029$ , that is the core size is much smaller and the whole configuration is dominated by the envelope solution only.

Figure 1 shows the plot of moment of inertia $I$ in units of $a^3$ ($`a' $ being the radius of the configuration) vs. $Q$ for assigned values of $u$ = 0.25 and 0.30.
Figures 2 - 7 sketch the variation of relative drag ($\Omega/\omega$) vs. $y$, that is the change in drag from centre ($y=0$) to the surface ($y=1$) of the configurations for $u$ values 0.25 and 0.30  considered in the present study and for different $Q$  values  from 0.001 to 1.3. It is seen that the the central drag $(\Omega/\omega)_0$ decreases with increasing $Q$ for both the values of $u = 0.25$ and 0.30. The surface drag  $(\Omega/\omega)_1$ remains almost same for 
all the values of $Q$ corresponding to the configurations with  $u$ values 0.25 and 0.30.

It is seen from Table 1 and 2 that for $Q$ values from 0.9 to 0.1, the core radius varies from about  83 to 29 \% of the total radius of the configuration and the density at the core-envelope boundary reaches from $2.5\times 10^{14}\rm {g cm}^{-3}$ to   $1.6\times 10^{15}\rm {g cm}^{-3}$. As a consequence, the fractional moment of inertia of the core component ($G_h$, the glitch healing parameter in the starquake glitch model) decreases from about 52 to 1.7 \% which is in excellent agreement with all the 11 measurements of $G_h$ for the Vela pulsar (see Table 1 of Crawford and Demiansky \cite{Ref3}). The range of the density at the core-envelope boundary mentioned above is also found  fully consistent with the range obtained  by Kalogera and Baym \cite{Ref6} who have used the modern EOSs (two variations of WFF  \cite{Ref18} EOS) between the said density range in the envelope region with a core given by extreme causal EOS, in context with the study of maximum mass of a  NS consistent with causality and dynamical stability. The present models are also found to be consistent with causality and dynamical stability.
\begin{table}
\caption{The Core-Envelope Boundary, $(b/a)$, Total Radius, $a$, Core Radius, $b$, Central  Energy-Density $E_0$, The ratios $E_0/E_b$ and $E_b/E_a$, Core Mass  $M_b/M_\odot$ and the glitch healing parameter,  $G_h = I_b/I_a$ , for the permitted value of $n = 1/2 (u = 0.25)$and  various allowed values of $Q$. The value of Surface Density, $E_a$, is assumed to be the average nuclear density, like Brecher \& Caporaso (\cite {Ref4}). For these values of $u$ and $E_a$ the total mass of the configuration correspond to a value of 2.263$M_\odot$.  }
\label{tab:1}       
\begin{tabular}{llllllllll}
\hline\noalign{\smallskip}
 &  &  & &  &  $ u = 0.25$ & &  &  &   \\
\hline\noalign{\smallskip}
$Q$ & $(b/a)$ &$ a$(km) &$ b$(km) & $8\pi E_0b^2$  & $E_0/E_b$ & $8\pi E_0a^2$  & $E_b/E_a$ & $M_b/M_\odot$ & $G_h = I_b/I_a$ \\
\noalign{\smallskip}\hline\noalign{\smallskip}
0.001 & 0.029 & 13.369 & 0.388 & 2.696  &6.284 & 3205.708  & 765.973 & 0.058 & 0.207E- 4 \\
0.100 & 0.288 & 13.369 & 3.850 & 2.589  & 5.857 & 31.214  & 8.002 & 0.565 & 0.019  \\
0.300 & 0.495 & 13.369 & 6.618 & 2.590  & 5.453 & 10.570  & 2.910 & 0.986 & 0.101 \\
0.600 & 0.693 & 13.369 & 9.265 & 2.654  & 5.008 & 5.526  & 1.657 & 1.442 & 0.295 \\
0.900 & 0.839 & 13.369 & 11.216 & 2.693 & 4.588 & 3.826 & 1.252 & 1.772 & 0.534 \\
1.300 & 0.993 & 13.369 & 13.275 & 2.744 & 4.139 & 2.783 & 1.010 & 2.240 & 0.974 \\
\noalign{\smallskip}\hline\noalign{\smallskip}
\end{tabular}
\end{table}

\begin{table}
\caption{The Core-Envelope Boundary, $(b/a)$, Total Radius, $a$, Core Radius, $b$, Central  Energy-Density $E_0$, The ratios $E_0/E_b$ and $E_b/E_a$, Core Mass  $M_b/M_\odot$ and the glitch healing parameter,  $G_h = I_b/I_a$ , for the permitted value of $n = 3/4 (u = 0.30)$and  various allowed values of $Q$. The value of Surface Density, $E_a$, is assumed to be the average nuclear density, like Brecher \& Caporaso (\cite {Ref4}). For these values of $u$ and $E_a$ the total mass of the configuration correspond to a value of 3.079$M_\odot$.}
\label{tab:2}
\begin{tabular}{llllllllll}
\hline\noalign{\smallskip}
 &  &  & &  &  $ u = 0.30$ & &  &  & \\
\hline\noalign{\smallskip}
$Q$ & $(b/a)$ &$ a$(km) &$ b$(km) & $8\pi E_0b^2$  & $E_0/E_b$ & $8\pi E_0a^2$  & $E_b/E_a$ & $M_b/M_\odot$ & $G_h = I_b/I_a$ \\
\noalign{\smallskip}\hline\noalign{\smallskip}
0.001 & 0.029 & 15.159 & 0.440 & 3.044  & 6.289 & 3619.501  & 671.562 & 0.075 & 0.188E- 4 \\
0.100 & 0.287 & 15.159 & 4.350 & 2.878  & 5.676 & 34.940  & 7.183 & 0.714 & 0.017  \\
0.300 & 0.493 & 15.159 & 7.472 & 2.960  & 5.267 & 12.179  & 2.698 & 1.282 & 0.093 \\
0.600 & 0.687 & 15.159 & 10.413 & 3.029 & 4.689 & 6.418 & 1.597 & 1.879 & 0.272 \\
0.900 & 0.830 & 15.159 & 12.580 & 3.116 & 4.263 & 4.523 & 1.238 & 2.392 & 0.518 \\
1.300 & 0.979 & 15.159 & 14.839 & 3.197 & 3.806 & 3.336 & 1.023 & 2.985 & 0.924 \\
\noalign{\smallskip}\hline\noalign{\smallskip}
\end{tabular}
\end{table}


\section{Summary of the Results}
The core-envelope models corresponding to compactness parameter, $u = 0.30$ and 0.25, obtained by Negi et al. \cite{Ref1}; Negi \cite{Ref2} are studied under slow rotation. The study shows that the models are capable of reproducing the all values of glitch healing parameter observed so far from the Vela pulsar on the basis of starquake glitch model. This result is in contrast to the earlier results of various studies which concluded that the starquake can not be a feasible mechanism for glitch generation in the Vela pulsar unless the Vela pulsar mass is unrealistic small. Since the matching of various parameter does not exist at the core-envelope boundary for the  value of $u \equiv M/a > 0.30$, the models predict the mass, $M$, surface redshift, $z_a$, and the moment of inertia, $I$, for the Vela pulsar as large as 3.079$M_\odot$, 0.581, and $ I_{\rm Vela,45} =6.9$ respectively for the $u$ value 0.30 and for an assigned value of the surface density, $E_a = 2 \times 10^{14}\rm {g cm}^{-3}$ \cite{Ref4} . However the values of various parameters less than those obtained  for the  Vela pulsar in the present study may be obtained likewise for the same values of glitch healing parameter, $G_h$, by assigning the $u$ values $< 0.25$ in the models described in   Negi et al. \cite{Ref1}; Negi \cite{Ref2}.

The study further indicates that the mass, moment of inertia and compactness of the Vela pulsar may be higher than that of the Crab pulsar if the starquake is found to be the feasible mechanism in the Vela pulsar together with the Crab. Since the core-envelope models considered in the present study may reproduce the identical results on mass, radius and density range of NS models  obtained by  Kalogera and Baym \cite{Ref6} by using various EOSs in the different region of the models, it follows that the simple analytical expressions relating radial coordinate directly to the pressure, energy-density and both of the metric parameters as used here may provide important information regarding internal structure of NSs.

\section {Conflict of Interest Statement:} The Author declares that there is no conflict of interest.

\begin{figure*}
  \includegraphics[width=0.75\textwidth]{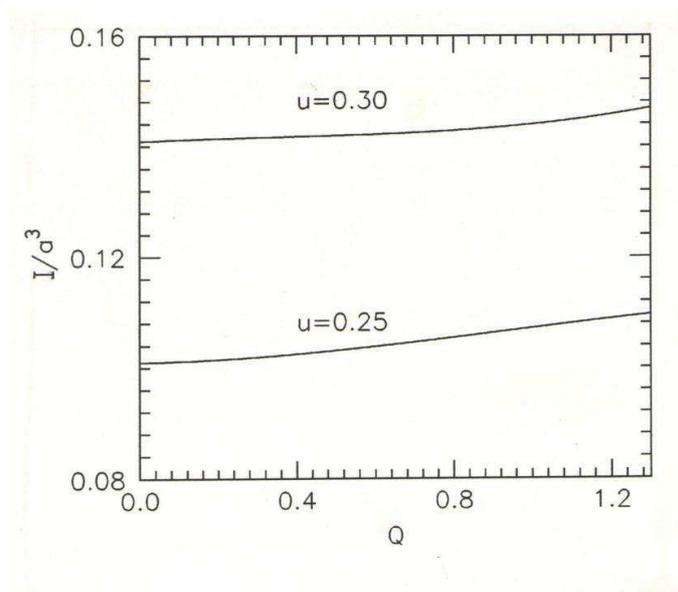}
\caption{Variation of $I/a^3$ with $Q$ for $u$ values 0.25 and 0.30.}
\label{fig:1}       
\end{figure*}

\begin{figure*}
  \includegraphics[width=0.75\textwidth]{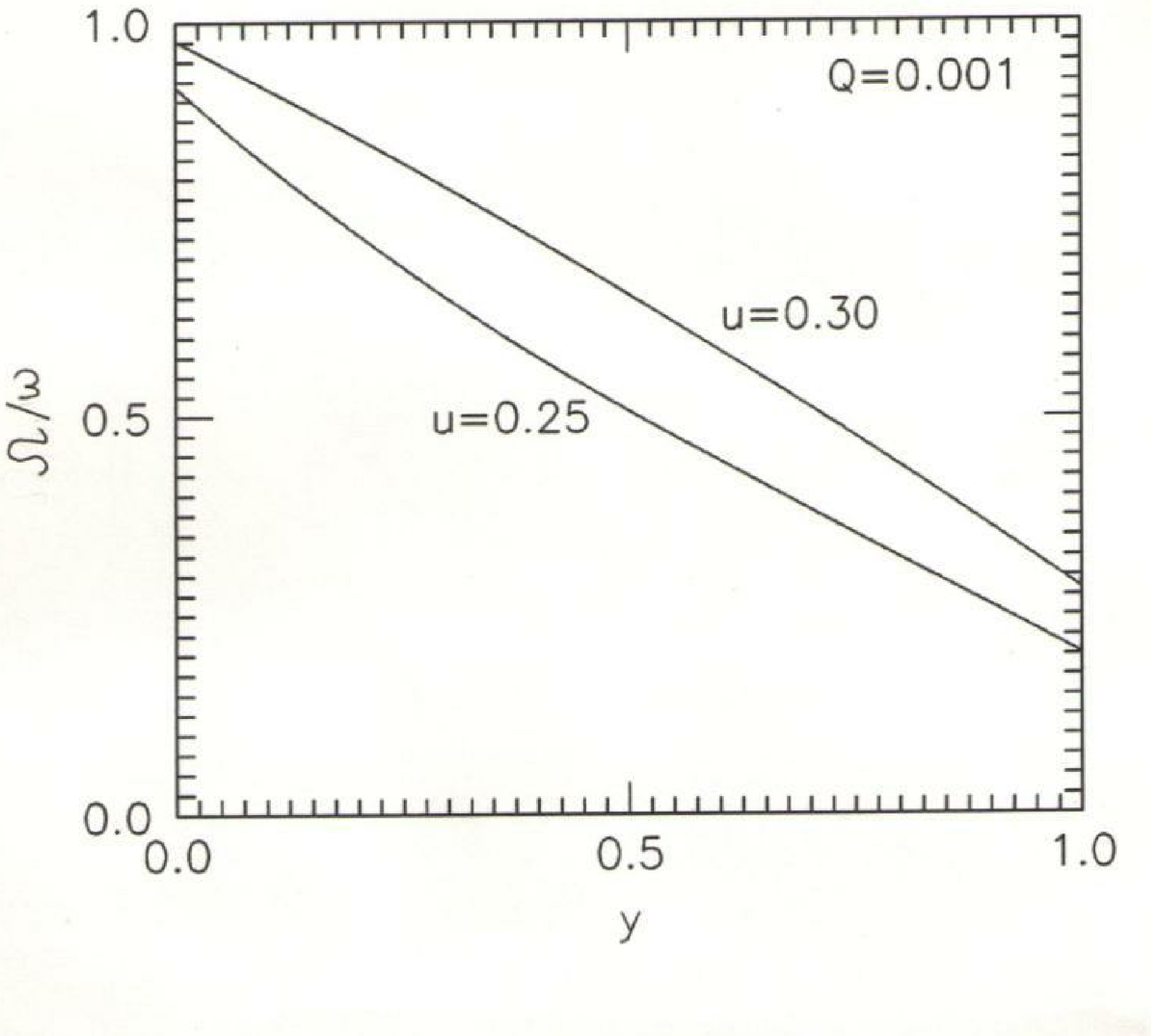}
\caption{ Variation of relative drag ($\Omega/\omega$) vs. $y$, that is the change in drag from centre ($y=0$) to the surface ($y=1$) of the structure for the value of $Q = 0.001$ and for the assigned $u$ values 0.25 and 0.30.}
\label{fig:2}       
\end{figure*}

\begin{figure*}
  \includegraphics[width=0.75\textwidth]{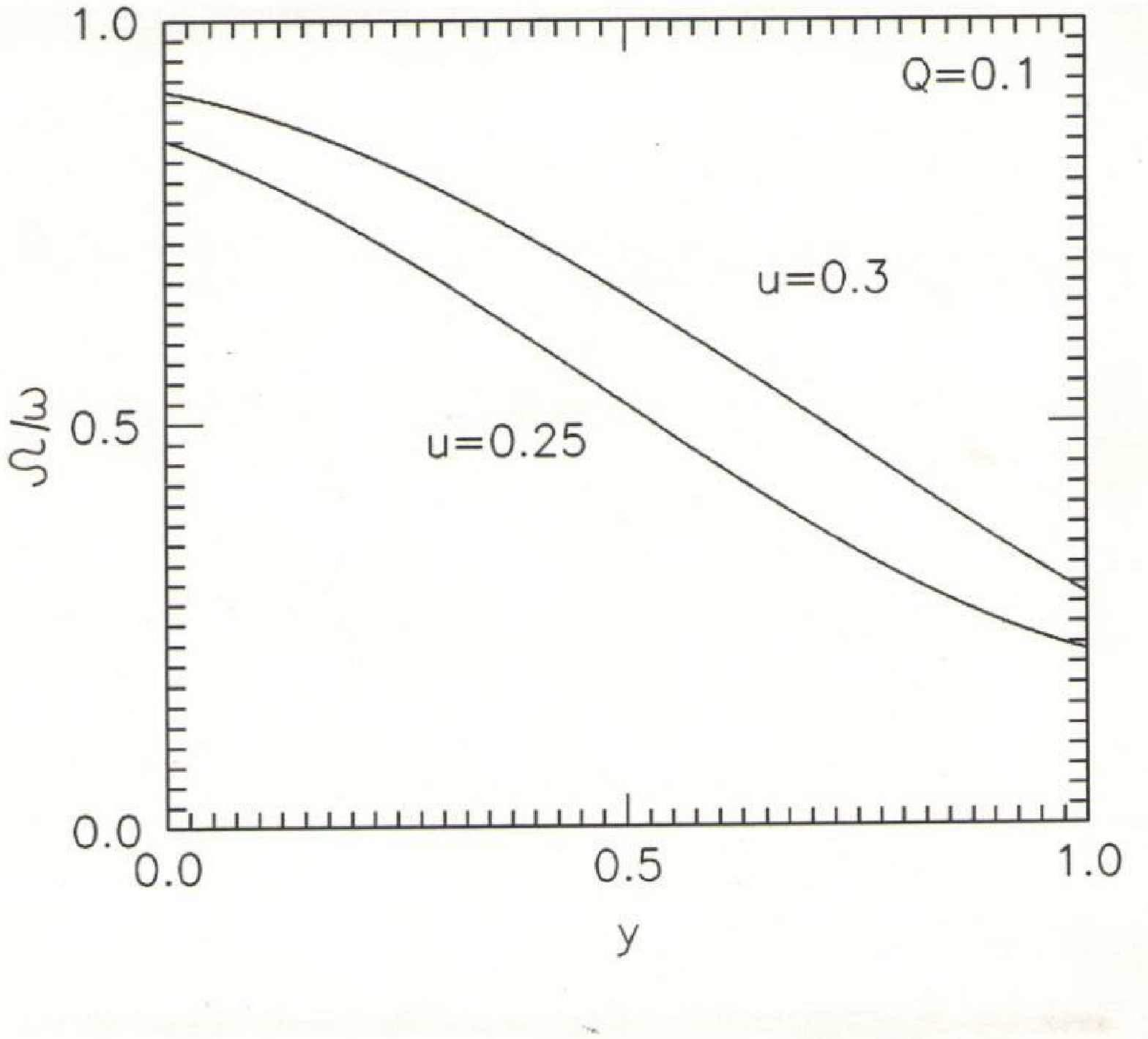}
\caption{ Variation of relative drag ($\Omega/\omega$) vs. $y$, that is the change in drag from centre ($y=0$) to the surface ($y=1$) of the structure for the value of $Q = 0.1$ and for the assigned $u$ values 0.25 and 0.30. }
\label{fig:3}       
\end{figure*}

\begin{figure*}
  \includegraphics[width=0.75\textwidth]{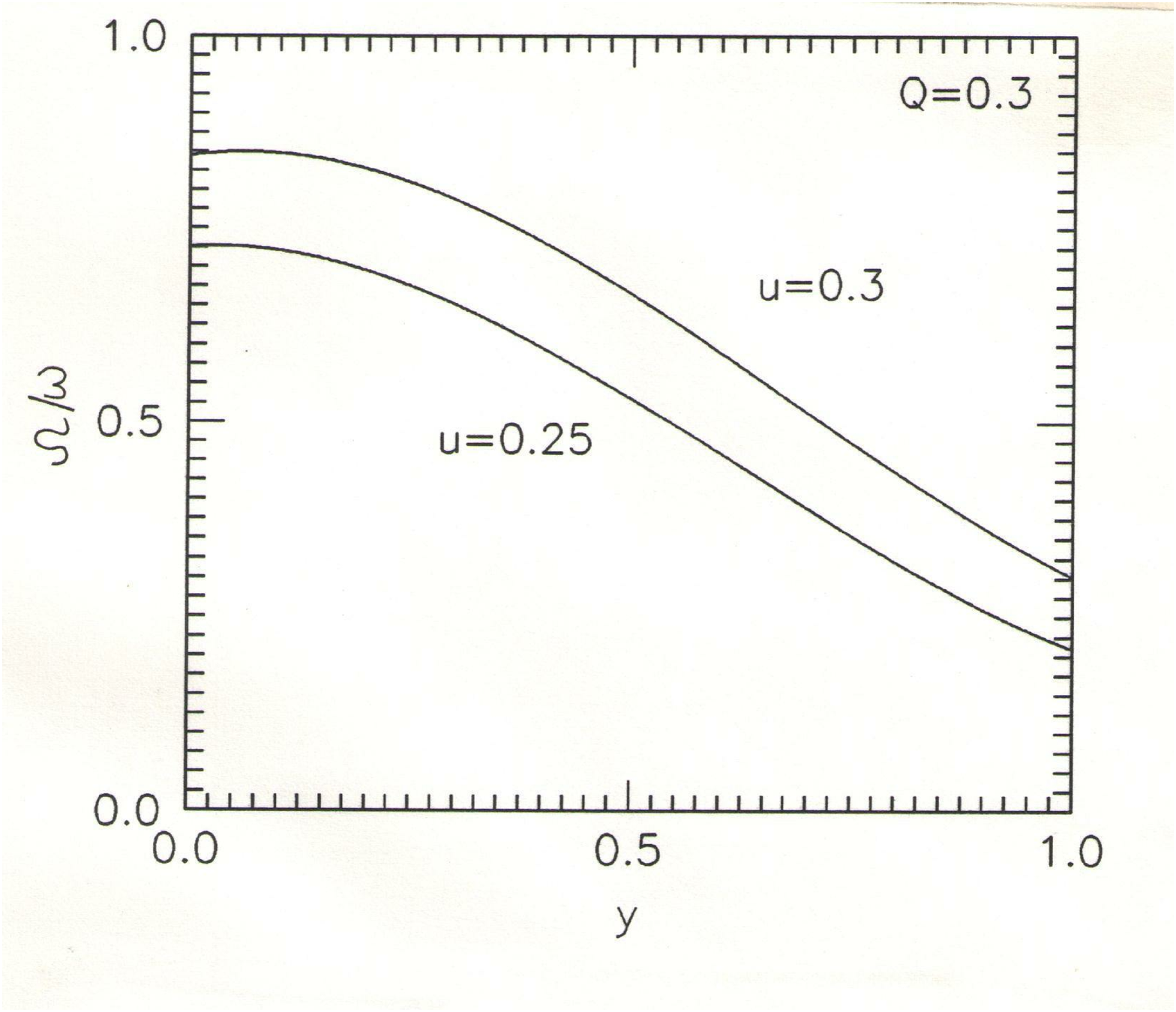}
\caption{Variation of relative drag ($\Omega/\omega$) vs. $y$, that is the change in drag from centre ($y=0$) to the surface ($y=1$) of the structure for the value of $Q = 0.3$ and for the assigned $u$ values 0.25 and 0.30. }
\label{fig:4}       
\end{figure*}

\begin{figure*}
  \includegraphics[width=0.75\textwidth]{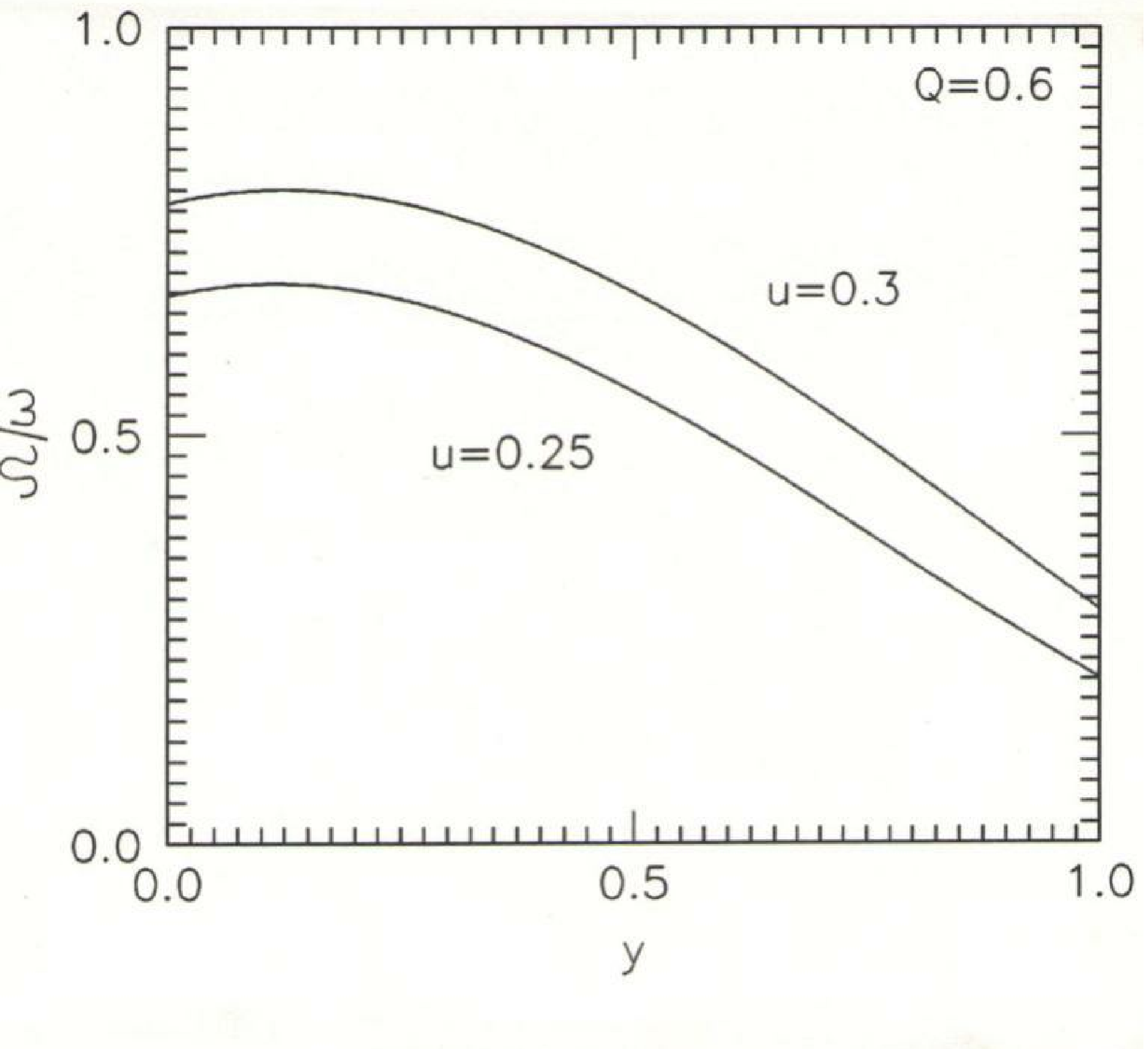}
\caption{Variation of relative drag ($\Omega/\omega$) vs. $y$, that is the change in drag from centre ($y=0$) to the surface ($y=1$) of the structure for the value of $Q = 0.6$ and for the assigned $u$ values 0.25 and 0.30. }
\label{fig:5}       
\end{figure*}

\begin{figure*}
  \includegraphics[width=0.75\textwidth]{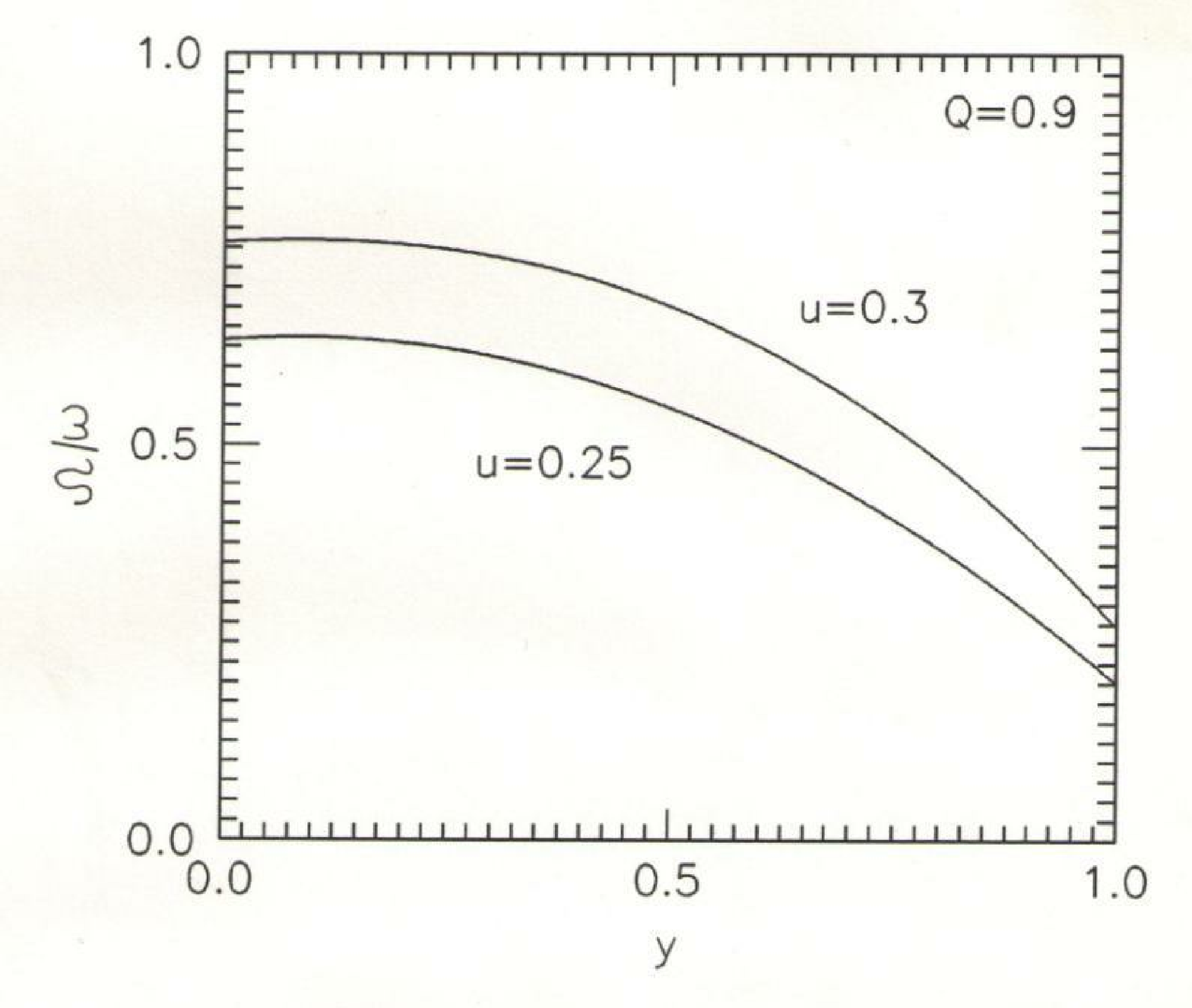}
\caption{Variation of relative drag ($\Omega/\omega$) vs. $y$, that is the change in drag from centre ($y=0$) to the surface ($y=1$) of the structure for the value of $Q = 0.9$ and for the assigned $u$ values 0.25 and 0.30. }
\label{fig:6}       
\end{figure*}

\begin{figure*}
  \includegraphics[width=0.75\textwidth]{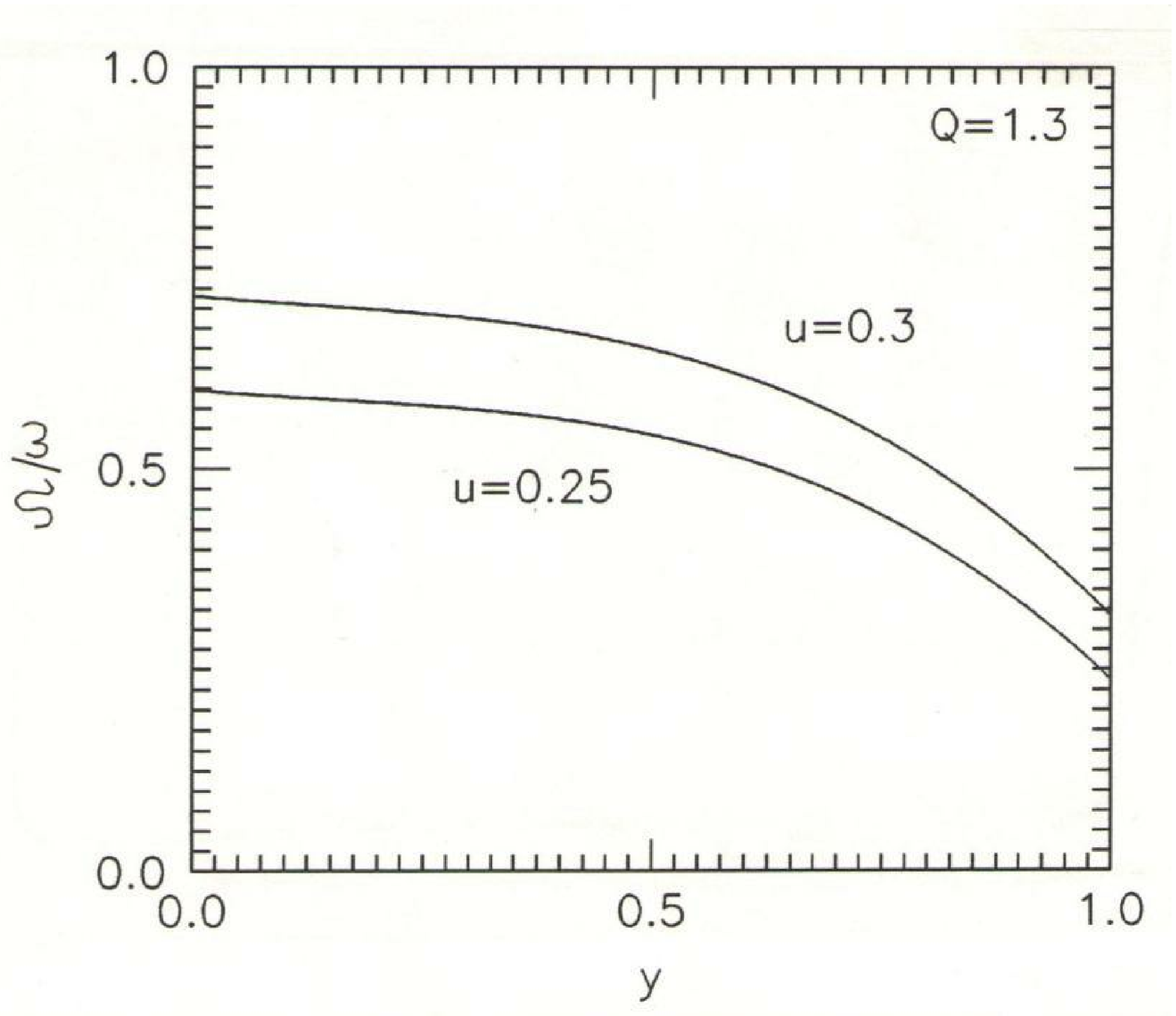}
\caption{Variation of relative drag ($\Omega/\omega$) vs. $y$, that is the change in drag from centre ($y=0$) to the surface ($y=1$) of the structure for the value of $Q = 1.3$ and for the assigned $u$ values 0.25 and 0.30. }
\label{fig:7}       
\end{figure*}

%




%
%


\end{document}